# Upconversion single photon detection near 2 μm

Guo-Liang Shentu,[1] Xiu-Xiu Xia,[1] Qi-Chao Sun,[1,2] Jason S. Pelc,[3] M. M. Fejer,[3] Qiang Zhang,[1,*] and Jian-Wei Pan[1]

[1]Shanghai Branch, National Laboratory for Physical Sciences at Microscale and Department of Modern Physics, University of Science and Technology of China, Shanghai, 201315, China
[2]Department of Physics, Shanghai Jiao Tong University, 800 Dongchuan Road, Shanghai, 200240, China
[3]Edward L. Ginzton Laboratory, Stanford University, Stanford, California 94305, USA
Corresponding author: qiangzh@ustc.edu.cn

We have demonstrated upconversion detection at the single photon level in the 2-μm spectral window using a pump wavelength near 1550nm, a periodically poled lithium niobate (PPLN) waveguide, and a volume Bragg grating (VBG) to reduce noise. We achieve a system photon detection efficiency of 10%, with a noise count rate of 24,500 counts per second, competitive with other 2-μm single photon detection technologies. This detector has potential applications in environmental gas monitoring, life science, and classical and quantum communication.

The detection of weak infrared signals at wavelengths near 2 μm is an enabling technology in fields such as environmental gas monitoring and life science [1-3], as many important chemical compounds have fundamental absorption bands located in this wavelength range due to vibrational or rotational transitions [3,4]. For example, $CO_2$ has an absorption band near 2 μm, the line strength of which is at least 70 times bigger than the absorption band located near 1.6 μm [5]. Therefore, the measurements of $CO_2$'s absorption at the 2 μm band can increase the altitude resolution comparing to the 1.6 μm band, and is more suitable for the lower troposphere [5]. However, traditional mid-infrared detection systems are not sensitive enough for those applications. In environmental gas monitoring, a detection sensitivity at least on the order of ~1 ppm (parts per million) is desired for $CO_2$, to define spatial gradients from which sources and sinks can be derived and quantified, as well as annual increase rate of its concentration [5]; and CO requires a detection sensitivity equivalent to ~100 ppb (parts per billion) [6]. Monitoring warming gases with detection sensitivity at these levels can help to study global warming and its influences for life on earth [7]. In life science, the on-line detection of several chemical compounds in exhaled human breath may help doctors with disease evaluation and treatment without invasion. For example, exhaled NO can be utilized as a biomarker for asthma [8], but it requires detection sensitivity at the level of ~1 ppb. Also, exhaled ammonia is a potential non-invasive marker of liver [9] and kidney function as well as peptic ulcer diseases.

Furthermore, hollow core-photonic band gap fibers (HC-PBGF) have been proposed to provide a 100 fold enhancement of the overall capacity of broadband core networks; recent results suggest that lowest propagation losses occur at a wavelength near 2 μm [10].

In summary, it can be seen that many fields can benefit from the single-photon detection at the 2-15 μm wavelength region. However, of the broadly employed InGaAs/InP avalanche photodiodes (APDs) to detect near infrared single photons, to the best of our knowledge no commercial detectors exist for wavelengths longer than 1.7 μm. Although commercial superconducting single-photon detectors (SSPDs) exist, reported detection efficiencies (DE) are approximately 3% at a temperature of 2 K [11]. Furthermore, SSPD's applications are limited due to for the need for bulky cryogenic cooling and exquisite temperature control. We have therefore investigated the application of frequency upconversion detectors [12-18] to the 2 μm band. In frequency upconversion, signal photons interact with a strong pump in a PPLN waveguide to produce converted photons in the visible to near-infrared region, which can be detected by a Silicon APD (SAPD) with a high DE and low dark count rates (DCRs).

In this work, we have demonstrated a room-temperature single-photon up-conversion detector at the 2μm band with a detection efficiency of approximately 10%.

We fabricated PPLN waveguides via the reverse proton exchange technique [19]; the waveguides are 52 mm long and are poled with a quasi-phase-matching (QPM) period of 19.6 μm. The waveguides were designed with single mode filters to match the SMF-28 single mode optical fiber, were fiber pigtailed at the input, and were antireflection coated to avoid interference effects and improve system throughput. A schematic of our experimental setup is shown in Fig. 1. A cw, single-frequency, tunable telecom-band external cavity diode laser (ECDL) and an erbium-doped fiber amplifier (EDFA) were utilized as the pump source for this experiment. The pump produced a maximum average power of approximately 1 W. The pump was filtered using a 1550-nm/1950-nm wavelength division multiplexer (WDM), a 1550-nm/980-nm WDM and a manually tunable optical filter to remove amplified spontaneous emission noise and other spurious output. A 1/99 beam splitter (BS) was utilized to monitor the pump power.

A single longitudinal mode Thulium fiber laser (TDFL) and amplifier (TDFA) were employed as the signal source near 1950 nm. Then a variable attenuator, five spliced 1/99 BSs and a thermal power meter with a sensitivity of 1 µW were utilized to control and monitor the input signal power. The total input photon number and time gate were set as 1 MHz and 1 ns, respectively. Therefore, the average photon number per time gate was $10^{-3}$. The signal was combined with the pump in a 1550nm/1950nm WDM. The polarization of both signal and pump fields were aligned to the TM mode of the PPLN waveguide using polarization controllers. Also, a Peltier cooler based temperature-control system was employed to keep the waveguide's temperature at 60 ℃, thereby maintaining the phase-matching condition. The signal, pump and SFG wavelengths are 1950 nm, 1550 nm and 864 nm, respectively.

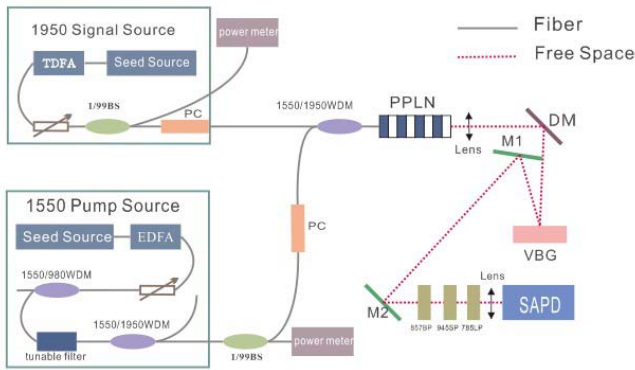

Fig. 1. Schematic of the up-conversion single photon detector at 2 µm. BS: beam splitter, M1-M2: mirrors.

The phase matching bandwidth of our waveguide was 0.3 nm, which was mainly decided by the length of the quasi phase matching gratings. To increase the bandwidth, one can reduce the length of the waveguide or fabricate a chirped one.

At the waveguide output, the sum-frequency generation （SFG) photons were separated from the pump by a dichroic mirror (DM) after collected by an AR-coated objective. A 945-nm short-pass filter (SPF), a 785-nm long-pass filter (LPF), and a 857-nm BPF were then used to block the second and higher order harmonic of the pump. Additionally, we used a volume Bragg grating (VBG) [18, 20] with a 95% reflection efficiency and a 0.05nm bandwidth to further suppress noise generated in the waveguide. Finally, the SFG photons were collected and detected by a SAPD, which had DE of approximately 45% at 860nm with the DCRs around 25 cps.

The main experimental results are shown in Fig. 2. We tuned the pump power by adjusting the variable attenuator and recorded DE and noise count rates at each tuned pump power point. The DE is calculated by dividing the number of detected counts after noise count rate subtraction and detector linearity correction by the number of signal photons before the WDM. When the pump power was set at 300 mW, the DE was 10.25%, with a noise count rate of 24,500 cps. DE data is fitted with a sine square function [14], while the noise count rate is fitted with a polynomial curve to guide the eye.

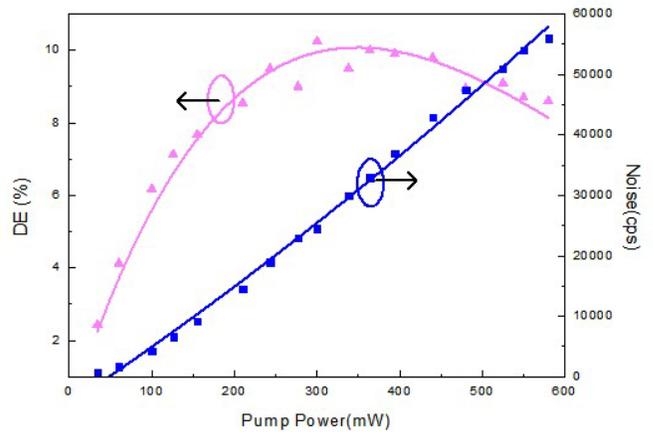

Fig. 2. The DE (purple) and noise count rate (blue) versus pump power. DE is fitted

The PPLN waveguide had a total fiber-to-output-facet throughput of −4.5 dB for the 2 µm signal. We measured the losses in the combination WDM and free space light path of -1 dB and -0.8 dB respectively. In combination with SAPD's DE, which is approximately 45%, the theoretical DE should be around 10%, which was consistent with our measurement.

In order to ascertain the conversion efficiency of the signal photons in the PPLN waveguide in our experiment, we measured the depletion [21] of the input signal as a function of the pump power by coupling the light exiting the waveguide into an OSA and comparing the observed signal levels when the pump is turned on versus off. A signal depletion level of -23.5 dB was observed, corresponding to an internal conversion efficiency of 99.6%. Our measurement of the signal depletion, in logarithmic units, is shown in Fig. 3. The experimental data fits the theory [21] very well.

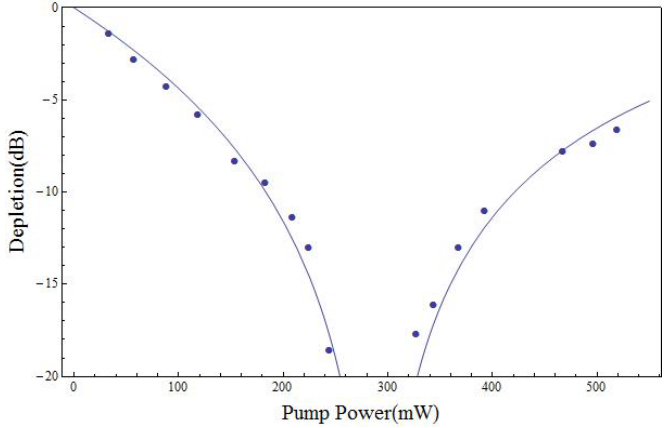

Fig. 3. Signal depletion versus pump power.

The noise is mainly due to spontaneous Raman scattering (SRS). SRS can produce photons either red- or blue-shifted from the pump as Stokes or anti-Stokes sidebands, which are generated with an intensity ratio determined by a Boltzmann factor owing to the thermal occupation of phonons [17]. The ratio of the rate $R_{aS}$ of anti-Stokes Raman photons for a medium pumped at $\omega_a$ to Stokes photons $R_S$ for a medium pumped at $\omega_b$, with $\omega_b - \omega_a = \Delta\omega$, is given by

$$\frac{R_{aS}}{R_s} = \left(\frac{\omega_b}{\omega_a}\right)^3 \exp\left[-\frac{h\Delta\omega}{2\pi kT}\right] \quad (1)$$

It was shown in [14] that for an up-conversion detector with $\lambda_p$ = 1.32μm and $\lambda_1$ = 1.55μm, the ratio of noise counts is consistent with this Boltzmann factor when the roles of the signal and pump are interchanged. In our experiment, the noise count rate was near 150 at the DE peak power when the pump and signal were interchanged, and the ratio calculated from equation (1) was appropriately 150. Therefore, a noise count rate of 24,500 cps at the DE peak is reasonable.

We have demonstrated the up-conversion detection near the 2-μm band with reasonable DCRs and a relatively high DE compared to existing single-photon detector at this wavelength, which has potential applications in many fields. To further improve the performance of this detector, we can employ long-wavelength pump [17] and cascade frequency conversion to reduce the noise as well as increasing the DE. A narrower band VBG could continue to reduce the noise. Note that the fiber coupling in the experiment was optimized for 1550 nm not 1950 nm. Therefore it is possible to redesign the mode filter to attain ~1 dB input coupling losses.

Note added.—We note that related experiment has been reported in Ref. [22].

### Acknowledge
The authors acknowledge Cheng-Zhi Peng and Yang Liu for their useful discussions. This work has been supported by the National Fundamental Research Program (under Grant No. 2011CB921300 and 2013CB336800), the NNSF of China, the CAS, and the Shandong Institute of Quantum Science & Technology Co., Ltd. J.S.P. and M.M.F. acknowledge the U.S. AFOSR for their support under Grant No. FA9550-09-1-0233.